\documentclass[aps,prl,reprint,showpacs,floatfix,superscriptaddress]{revtex4-1}
\usepackage{amsmath}
\usepackage{amssymb}
\usepackage{graphicx}
\usepackage{epsfig}

\newcommand{\fpi}{\mbox{$F_\pi$}}
\newcommand{\qsq}{\mbox{$Q^2$}}
\newcommand{\sigl}{\mbox{$\sigma_L$}}
\newcommand{\sigt}{\mbox{$\sigma_T$}}

\newcommand{\gevsq}{\mbox{GeV$^2$}}

\graphicspath{{/home/huberg/r2d2/piminus/paper/}}
\DeclareGraphicsExtensions{ .eps , .eps.gz ,
                            .ps  , .ps.gz }

\begin{document}
\verb| |\\  

\title{
Separated Response Function Ratios in Exclusive, Forward $\bf\pi^{\pm}$
Electroproduction}

\author{G.M.~Huber}
\affiliation{University of Regina, Regina, Saskatchewan S4S 0A2, Canada}
\author{H.P.~Blok}
\affiliation{VU university, NL-1081 HV Amsterdam, The Netherlands}
\affiliation{NIKHEF, Postbus 41882, NL-1009 DB Amsterdam, The Netherlands}
\author {C.~Butuceanu}
\affiliation{University of Regina, Regina, Saskatchewan S4S 0A2, Canada}
\author{D.~Gaskell}
\affiliation{Thomas Jefferson National Accelerator Facility,
 Newport News, Virginia 23606}
\author{T.~Horn}
\affiliation{Catholic University of America, Washington, DC 20064}
\author{D.J.~Mack}
\affiliation{Thomas Jefferson National Accelerator Facility,
 Newport News, Virginia 23606}
\author{D.~Abbott}
\affiliation{Thomas Jefferson National Accelerator Facility,
 Newport News, Virginia 23606}
\author{K.~Aniol}
\affiliation{California State University Los Angeles, Los Angeles, California
  90032}
\author{H.~Anklin}
\affiliation{Florida International University, Miami, Florida 33119}
\affiliation{Thomas Jefferson National Accelerator Facility,
 Newport News, Virginia 23606}
\author{C.~Armstrong}
\affiliation{College of William and Mary, Williamsburg, Virginia 23187}
\author{J.~Arrington}
\affiliation{Physics Division, Argonne National Laboratory, Argonne, Illinois
  60439}
\author{K.~Assamagan}
\affiliation{Hampton University, Hampton, Virginia 23668}
\author{S.~Avery}
\affiliation{Hampton University, Hampton, Virginia 23668}
\author{O.K.~Baker}
\affiliation{Hampton University, Hampton, Virginia 23668}
\affiliation{Thomas Jefferson National Accelerator Facility,
 Newport News, Virginia 23606}
\author{B.~Barrett}
\affiliation{Saint Mary's University, Halifax, Nova Scotia B3H 3C3 Canada}
\author{E.J.~Beise}
\affiliation{University of Maryland, College Park, Maryland 20742}
\author{C.~Bochna}
\affiliation{University of Illinois, Champaign, Illinois 61801}
\author{W.~Boeglin}
\affiliation{Florida International University, Miami, Florida 33119}
\author{E.J.~Brash}
\affiliation{University of Regina, Regina, Saskatchewan S4S 0A2, Canada}
\author{H.~Breuer}
\affiliation{University of Maryland, College Park, Maryland 20742}
\author{C.C.~Chang}
\affiliation{University of Maryland, College Park, Maryland 20742}
\author{N.~Chant}
\affiliation{University of Maryland, College Park, Maryland 20742}
\author{M.E.~Christy}
\affiliation{Hampton University, Hampton, Virginia 23668}
\author{J.~Dunne}
\affiliation{Thomas Jefferson National Accelerator Facility,
 Newport News, Virginia 23606}
\author{T.~Eden}
\affiliation{Thomas Jefferson National Accelerator Facility,
 Newport News, Virginia 23606}
\affiliation{Norfolk State University, Norfolk, Virginia 23504}
\author{R.~Ent}
\affiliation{Thomas Jefferson National Accelerator Facility,
 Newport News, Virginia 23606}
\author{H.~Fenker}
\affiliation{Thomas Jefferson National Accelerator Facility,
 Newport News, Virginia 23606}
\author{E.F.~Gibson}
\affiliation{California State University, Sacramento, California 95819}
\author{R.~Gilman}
\affiliation{Rutgers, The State University of New Jersey, Piscataway, New
Jersey 08854}
\affiliation{Thomas Jefferson National Accelerator Facility,
 Newport News, Virginia 23606}
\author{K.~Gustafsson}
\affiliation{University of Maryland, College Park, Maryland 20742}
\author{W.~Hinton}
\affiliation{Hampton University, Hampton, Virginia 23668}
\author{R.J.~Holt}
\affiliation{Physics Division, Argonne National Laboratory, Argonne, Illinois
  60439}
\author{H.~Jackson}
\affiliation{Physics Division, Argonne National Laboratory, Argonne, Illinois
  60439}
\author{S.~Jin}
\affiliation{Kyungpook National University, Daegu, 702-701, Republic of Korea}
\author{M.K.~Jones}
\affiliation{College of William and Mary, Williamsburg, Virginia 23187}
\author{C.E.~Keppel}
\affiliation{Hampton University, Hampton, Virginia 23668}
\affiliation{Thomas Jefferson National Accelerator Facility,
 Newport News, Virginia 23606}
\author{P.H.~Kim}
\affiliation{Kyungpook National University, Daegu, 702-701, Republic of Korea}
\author{W.~Kim}
\affiliation{Kyungpook National University, Daegu, 702-701, Republic of Korea}
\author{P.M.~King}
\affiliation{University of Maryland, College Park, Maryland 20742}
\author{A.~Klein}
\affiliation{Old Dominion University, Norfolk, Virginia 23529}
\author{D.~Koltenuk}
\affiliation{University of Pennsylvania, Philadelphia, Pennsylvania 19104}
\author{V.~Kovaltchouk}
\affiliation{University of Regina, Regina, Saskatchewan S4S 0A2, Canada}
\author{M.~Liang}
\affiliation{Thomas Jefferson National Accelerator Facility,
 Newport News, Virginia 23606}
\author{J.~Liu}
\affiliation{University of Maryland, College Park, Maryland 20742}
\author{G.J.~Lolos}
\affiliation{University of Regina, Regina, Saskatchewan S4S 0A2, Canada}
\author{A.~Lung}
\affiliation{Thomas Jefferson National Accelerator Facility,
 Newport News, Virginia 23606}
\author{D.J.~Margaziotis}
\affiliation{California State University Los Angeles, Los Angeles, California
  90032}
\author{P.~Markowitz}
\affiliation{Florida International University, Miami, Florida 33119}
\author{A.~Matsumura}
\affiliation{Tohoku University, Sendai, Japan}
\author{D.~McKee}
\affiliation{New Mexico State University, Las Cruces, New Mexico 88003-8001}
\author{D.~Meekins}
\affiliation{Thomas Jefferson National Accelerator Facility,
 Newport News, Virginia 23606}
\author{J.~Mitchell}
\affiliation{Thomas Jefferson National Accelerator Facility,
 Newport News, Virginia 23606}
\author{T.~Miyoshi}
\affiliation{Tohoku University, Sendai, Japan}
\author{H.~Mkrtchyan}
\affiliation{A.I. Alikhanyan National Science Laboratory, Yerevan 0036,
  Armenia}
\author{B.~Mueller}
\affiliation{Physics Division, Argonne National Laboratory, Argonne, Illinois
  60439}
\author{G.~Niculescu}
\affiliation{James Madison University, Harrisonburg, Virginia 22807}
\author{I.~Niculescu}
\affiliation{James Madison University, Harrisonburg, Virginia 22807}
\author{Y.~Okayasu}
\affiliation{Tohoku University, Sendai, Japan}
\author{L.~Pentchev}
\affiliation{College of William and Mary, Williamsburg, Virginia 23187}
\author{C.~Perdrisat}
\affiliation{College of William and Mary, Williamsburg, Virginia 23187}
\author{D.~Pitz}
\affiliation{DAPNIA/SPhN, CEA/Saclay, F-91191 Gif-sur-Yvette, France}
\author{D.~Potterveld}
\affiliation{Physics Division, Argonne National Laboratory, Argonne, Illinois
  60439}
\author{V.~Punjabi}
\affiliation{Norfolk State University, Norfolk, Virginia 23504}
\author{L.M.~Qin}
\affiliation{Old Dominion University, Norfolk, Virginia 23529}
\author{P.E.~Reimer}
\affiliation{Physics Division, Argonne National Laboratory, Argonne, Illinois
  60439}
\author{J.~Reinhold}
\affiliation{Florida International University, Miami, Florida 33119}
\author{J.~Roche}
\affiliation{Thomas Jefferson National Accelerator Facility,
 Newport News, Virginia 23606}
\author{P.G.~Roos}
\affiliation{University of Maryland, College Park, Maryland 20742}
\author{A.~Sarty}
\affiliation{Saint Mary's University, Halifax, Nova Scotia B3H 3C3 Canada}
\author{I.K.~Shin}
\affiliation{Kyungpook National University, Daegu, 702-701, Republic of Korea}
\author{G.R.~Smith}
\affiliation{Thomas Jefferson National Accelerator Facility,
 Newport News, Virginia 23606}
\author{S.~Stepanyan}
\affiliation{A.I. Alikhanyan National Science Laboratory, Yerevan 0036,
  Armenia}
\author{L.G.~Tang}
\affiliation{Hampton University, Hampton, Virginia 23668}
\affiliation{Thomas Jefferson National Accelerator Facility,
 Newport News, Virginia 23606}
\author{V.~Tadevosyan}
\affiliation{A.I. Alikhanyan National Science Laboratory, Yerevan 0036,
  Armenia}
\author{V.~Tvaskis}
\affiliation{VU university, NL-1081 HV Amsterdam, The Netherlands}
\affiliation{NIKHEF, Postbus 41882, NL-1009 DB Amsterdam, The Netherlands}
\author{R.L.J.~van~der~Meer}
\affiliation{University of Regina, Regina, Saskatchewan S4S 0A2, Canada}
\author{K.~Vansyoc}
\affiliation{Old Dominion University, Norfolk, Virginia 23529}
\author{D.~Van~Westrum}
\affiliation{University of Colorado, Boulder, Colorado 80309}
\author{S.~Vidakovic}
\affiliation{University of Regina, Regina, Saskatchewan S4S 0A2, Canada}
\author{J.~Volmer}
\affiliation{VU university, NL-1081 HV Amsterdam, The Netherlands}
\affiliation{DESY, Hamburg, Germany}
\author{W.~Vulcan}
\affiliation{Thomas Jefferson National Accelerator Facility,
 Newport News, Virginia 23606}
\author{G.~Warren}
\affiliation{Thomas Jefferson National Accelerator Facility,
 Newport News, Virginia 23606}
\author{S.A.~Wood}
\affiliation{Thomas Jefferson National Accelerator Facility,
 Newport News, Virginia 23606}
\author{C.~Xu}
\affiliation{University of Regina, Regina, Saskatchewan S4S 0A2, Canada}
\author{C.~Yan}
\affiliation{Thomas Jefferson National Accelerator Facility,
 Newport News, Virginia 23606}
\author{W.-X.~Zhao}
\affiliation{Massachusetts Institute of Technology, Cambridge, 
Massachusetts 02139}
\author{X.~Zheng}
\affiliation{Physics Division, Argonne National Laboratory, Argonne, Illinois
  60439}
\author{B.~Zihlmann}
\affiliation{University of Virginia, Charlottesville, Virginia 22901}
\affiliation{Thomas Jefferson National Accelerator Facility,
 Newport News, Virginia 23606}
\collaboration{The Jefferson Lab \fpi\ Collaboration}
\noaffiliation

\date{\today}
 
\begin{abstract}
The study of exclusive $\pi^{\pm}$ electroproduction on the nucleon, including
separation of the various structure functions, is of interest for a number of
reasons.  The ratio $R_L=\sigma_L^{\pi^-}/\sigma_L^{\pi^+}$ is sensitive to
isoscalar contamination to the dominant isovector pion exchange amplitude,
which is the basis for the determination of the charged pion form factor from
electroproduction data.  A change in the value of
$R_T=\sigma_T^{\pi^-}/\sigma_T^{\pi^+}$ from unity at small $-t$, to 1/4 at
large $-t$, would suggest a transition from coupling to a (virtual) pion to
coupling to individual quarks.  Furthermore, the mentioned ratios may show an
earlier approach to pQCD than the individual cross sections.  We have performed
the first complete separation of the four unpolarized electromagnetic structure
functions above the dominant resonances in forward, exclusive $\pi^{\pm}$
electroproduction on the deuteron at central \qsq\ values of 0.6, 1.0, 1.6
\gevsq\ at $W$=1.95 GeV, and $Q^2=2.45$ \gevsq\ at $W$=2.22 GeV.  Here, we
present the $L$ and $T$ cross sections, with emphasis on $R_L$ and $R_T$, and
compare them with theoretical calculations.  Results for the separated ratio
$R_L$ indicate dominance of the pion-pole diagram at low $-t$, while results
for $R_T$ are consistent with a transition between pion knockout and quark
knockout mechanisms.
\end{abstract}


\maketitle

Measurements of exclusive meson production are a useful tool in the study of
hadronic structure.  Through these studies, one can discern the relevant
degrees of freedom at different distance scales.  In contrast to inclusive
$(e,e')$ or photoproduction measurements, the transverse momentum (size) of a
scattering constituent and the resolution at which it is probed can be varied
independently.  Exclusive {\em forward pion} electroproduction is especially
interesting, because by detecting the charge of the pion, even the flavor of
the interacting constituents can be tagged.  Finally, {\em ratios} of separated
response functions can be formed for which nonperturbative corrections may
partially cancel, yielding insight into soft-hard factorization at the modest
photon virtuality, \qsq, to which exclusive measurements will be limited for
the foreseeable future.

The longitudinal response in exclusive charged pion electroproduction has
several important applications.  At low Mandelstam variable $-t$, it can be
related to the charged pion form factor, $F_{\pi}(Q^2)$, \cite{Huber08} which
is used to test non-perturbative models of this ``positronium'' of light quark
QCD.  In order to reliably extract $F_{\pi}$ from electroproduction data, the
isovector $t$-pole process should be dominant in the kinematic region under
study.  This dominance can be studied experimentally through the ratio of
longitudinal $\gamma^{*}_L n \to \pi^- p$ and $\gamma^*_L p \to \pi^+ n$ cross
sections.  If the photon possessed definite isospin, exclusive $\pi^-$
production on the neutron and $\pi^+$ production on the proton would be related
to each other by simple isospin rotation and the cross sections would be equal
\cite{boyarski68}.  A departure from $R_L\equiv\sigma_L^{\pi^-}
/\sigma_L^{\pi^+}= \frac{|A_V-A_S|^2}{|A_V+A_S|^2}=1$, where $A_S$ and $A_V$
are the respective isoscalar and isovector photon amplitudes, would indicate
the presence of isoscalar backgrounds arising from mechanisms such as $\rho$
meson exchange \cite{VGL1} or perturbative contributions due to transverse
quark momentum \cite{Milana}.  Such physics backgrounds may be expected to be
larger at higher $-t$ (due to the drop-off of the pion pole) or non-forward
kinematics (due to angular momentum conservation).  Because previous data are
unseparated \cite{Brauel1}, no firm conclusions about possible deviations of
$R_L$ from unity were possible.

In the limit of small $-t$, where the photon is expected to couple to the
charge of the pion, the transverse ratio
$R_T\equiv\sigma_T^{\pi^-}/\sigma_T^{\pi^+}$ is expected to be near unity.
With increasing $-t$, the photon starts to probe quarks rather than pions, and
the charge of the produced pion acts as a tag on the flavor of the
participating constituent.  Applying isospin decomposition and charge symmetry
invariance to $s$-channel knockout of valence quarks in the hard-scattering
regime, Nachtmann \cite{nachtmann} predicted the exclusive electroproduction
$\pi^-/\pi^+$ ratio at sufficiently large $-t$ to be $\frac{\gamma^*_T
n\rightarrow\pi^-p}{\gamma^*_T
p\rightarrow\pi^+n}=\Bigl(\frac{e_d}{e_u}\Bigr)^2=\frac{1}{4}$.  Previous
unseparated $\pi^-/\pi^+$ data \cite{Brauel1} trend to a ratio of 1/4 for
$|t|>0.6$ \gevsq, but with relatively large uncertainties.

In the transition region between low $-t$ (where a description of hadronic
degrees of freedom in terms of effective hadronic Lagrangians is valid) and
large $-t$ (where the degrees of freedom are quarks and gluons), $t$-channel
exchange of a few Regge trajectories permits an efficient description of the
energy dependence and the forward angular distribution of many real- and
virtual-photon-induced reactions.  The VGL Regge model \cite{VGL,van98} has
provided a good and consistent description of a wide variety of $\pi^{\pm}$
photo- and electroproduction data above the resonance region.  However, the
model has consistently failed to provide a good description of $p(e,e'\pi^+)n$
\sigt\ data \cite{Blok08}.  The VGL Regge model was recently extended
\cite{kaskulov, vrancx} by the addition of a hard deep inelastic scattering
(DIS) process of virtual-photons off nucleons.  The DIS process dominates the
transverse response at moderate and high \qsq, providing a better description
of \sigt.

Exclusive $\pi^{\pm}$ electroproduction has also been calculated in the handbag
framework, where only one parton participates in the hard subprocess, and the
soft physics is encoded in generalized parton distributions (GPDs).
Pseudoscalar meson production, such as $\sigt$ in exclusive $\pi^{\pm}$
electroproduction which is not dominated by the pion pole term, has been
identified as being especially sensitive to the chiral-odd transverse GPDs
\cite{ahmad,gk10}.  The model of Refs. \cite{gk10,gk13} uses a modified
perturbative approach based on GPDs, incorporating the full pion
electromagnetic form factor and substantial contributions from the twist-3
transversity GPD, $H_T$.

We have performed a complete $L$/$T$/$LT$/$TT$ separation in exclusive forward
$\pi^{\pm}$ electroproduction from deuterium.  Here, we present the $L$ and $T$
cross sections, with emphasis on $R_L$ and $R_T$ in order to better understand
the dynamics of this fundamental inelastic process; the $LT$ and $TT$
interference cross sections will be presented in a future work.  Because there
are no practical free neutron targets, the $^2$H$(e,e'\pi^{\pm})NN_s$ reactions
(where $N_s$ denotes the spectator nucleon) were used.  In $\pi^-/\pi^+$
ratios, the corrections for nuclear binding and rescattering largely cancel.

The data were obtained in Hall C at the Thomas Jefferson National Accelerator
Facility (JLab) as part of the two pion form factor experiments presented in
detail in Ref. \cite{Blok08}.  Except where noted, the experimental details and
data analysis techniques are as presented in Ref. \cite{Blok08} for the
$^1$H$(e,e'\pi^+)n$ data.  Charged $\pi^{\pm}$ were detected in the High
Momentum Spectrometer (HMS) while the scattered electrons were detected in the
Short Orbit Spectrometer (SOS).  Given the kinematic constraints imposed by the
available electron beam energies and the properties of the HMS and SOS magnetic
spectrometers, deuterium data were acquired in the first experiment for nominal
$(Q^2$, $W$, $\Delta\epsilon)$ settings of $(0.60, 1.95, 0.37)$, $(1.00, 1.95,
0.32)$, $(1.60,1.95, 0.36)$, and in the second experiment of $(2.45, 2.22,
0.27)$.  The value $W$=1.95 GeV used in the first experiment is high enough to
suppress most $s$-channel baryon resonance backgrounds, but this suppression
should be even more effective in the second experiment.  For each \qsq\
setting, the electron spectrometer angle and momentum, as well as the pion
spectrometer momentum, were kept fixed.  To attain full coverage in $\phi$, in
most cases additional data were taken with the pion spectrometer at a slightly
smaller and at a larger angle than the $\vec{q}$-vector direction for the high
$\epsilon$ settings.  At low $\epsilon$, only the larger angle setting was
possible.  The HMS magnetic polarity was reversed between $\pi^+$ and $\pi^-$
running, with the quadrupoles and dipole magnets cycled according to a standard
procedure.  Kinematic offsets in spectrometer angle and momentum, as well as in
beam energy, were previously determined using elastic $e^-p$ coincidence data
taken during the same run, and the reproducibility of the optics checked
\cite{Blok08}.

The potential contamination by electrons when the pion spectrometer is set to
negative polarity, and by protons when it is set to positive polarity,
introduces some differences in the $\pi^{\pm}$ data analyses which were
carefully examined.  For most negative HMS polarity runs, electrons were
rejected at the trigger level by a gas \u{C}erenkov detector containing ${\rm
C}_4{\rm F}_{10}$.  The beam current was significantly reduced during $\pi^-$
running to minimize the inefficiency due to electrons passing through the gas
\u{C}erenkov within $\approx 100$ ns after a pion has traversed the detector,
causing the pion to be misidentified as an electron.  A \u{C}erenkov blocking
correction (1-15\%) was applied to the $\pi^-$ data using the measured electron
rates combined with the effective time window of the gas \u{C}erenkov ADC, the
latter determined from data where the \u{C}erenkov was not in the trigger.  A
cut on particle speed ($v/c>0.95$), calculated from the time-of-flight
difference between two scintillator planes in the HMS detector stack, was used
to separate $\pi^+$ from protons.  Additionally in the second experiment, an
aerogel \u{C}erenkov detector was used to separate protons and $\pi^+$ for
central momenta above 3 GeV/$c$.  A correction for the number of pions lost due
to pion nuclear interactions and true absorption in the HMS exit window and
detector stack of 4.5-6\% was applied.  For further details, see
Ref. \cite{Blok08}.

Because the $\pi^-$ data are typically taken at higher HMS detector rates than
the $\pi^+$ data, a good understanding of rate-dependent efficiency corrections
was required.  An improved high rate tracking algorithm was implemented,
resulting in high rate tracking inefficiencies of 2-9\% for HMS rates up to 1.4
MHz.  Liquid deuterium target boiling corrections of 4.7\%/100 $\mu$A were
determined for the horizontal-flow target used in the first experiment.  The
vertical-flow target and improved beam raster used in the second experiment
resulted in a negligible boiling correction for those data.  The experimental
yields were also corrected for dead time (1-11\%).

\begin{figure}[h]
\begin{center}
\includegraphics[width=3.25in]{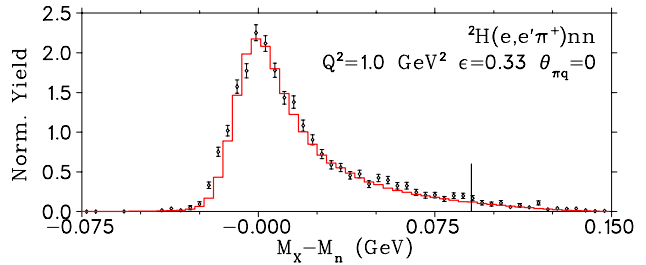}
\end{center}
\caption{(Color online) Missing mass of the undetected nucleon calculated as
  quasi-free pion electroproduction for a representative $\pi^+$ setting.  
  The diamonds are experimental data, and the red line is the quasi-free Monte
  Carlo simulation.  The vertical line indicates the $M_X$ cut upper limit.}
\label{fig:MMplot}
\end{figure}

Kinematic quantities such as $t$ and missing mass $M_X$ were reconstructed as
quasi-free pion electroproduction, $\gamma^* N \rightarrow \pi^{\pm} N'$, where
the virtual-photon interacts with a nucleon at rest.  The former is calculated
using $t=(p_{\rm target}-p_{\rm recoil})^2$, which can differ from
$(p_{\gamma}-p_{\pi})^2$ due to Fermi motion and radiation.  Missing mass cuts
were then applied to select the exclusive final state (Fig. \ref{fig:MMplot}).
Because of Fermi motion in the deuteron, this cut is taken wider than for a
hydrogen target.  Real and random coincidences were isolated with a coincidence
time cut of $\pm 1$ ns.  Background from aluminum target cell walls (2-4\% of
the yield) and random coincidences ($\sim 1\%$) were subtracted from the
charge-normalized yields on a bin by bin basis.

The virtual-photon cross section can be expressed in terms of contributions
from transversely and longitudinally polarized photons, and interference terms,
\begin{eqnarray}
\label{eqn:unsep}
  2\pi \frac{d^2 \sigma}{dt d\phi} & = & \frac{d \sigma_T}{dt} +
  \epsilon \frac{d \sigma_L}{dt} + \sqrt{2 \epsilon (1 + \epsilon)}
  \frac{d \sigma_{LT}}{dt} \cos \phi \\ \nonumber & + & \epsilon \frac{d
  \sigma_{TT}}{dt} \cos 2 \phi.
\end{eqnarray}
Here,
$\epsilon=\left(1+2\frac{|\vec{q}|^2}{Q^2}\tan^2\frac{\theta}{2}\right)^{-1}$
is the virtual-photon polarization, where $\vec{q}$ is the three-momentum
transferred to the quasi-free nucleon, $\theta$ is the electron scattering
angle, and $\phi$ is the azimuthal angle between the scattering and the
reaction plane.

\begin{figure}
\begin{center}
\includegraphics[width=3.5in]{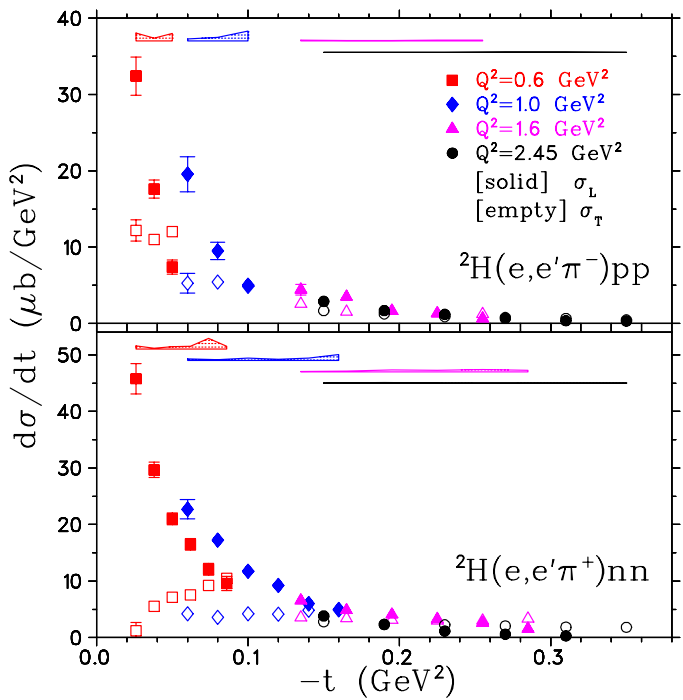}
\caption{(Color online) Separated exclusive $\pi^{\pm}$ electroproduction cross
  sections from deuterium.  Because the data were
  taken at different values of $\overline{W}$, all cross sections were scaled
  to a value of $W=2.0$ GeV according to $1/(W^2-M^2)$.
  The error bars indicate statistical and uncorrelated systematic uncertainties
  in both $\epsilon$ and $-t$, combined in quadrature.
  The shaded error bands indicate the model-dependence of $\sigl$.
  The $\sigt$ model-dependence (not shown) is smaller.
\label{fig:xsec}
}
\end{center}
\end{figure}

For each charge state, the data for $d^2\sigma/dtd\phi$ were binned in $t$ and
$\phi$ and the individual components in Eqn.~\ref{eqn:unsep} determined from a
simultaneous fit to the $\phi$ dependence of the measured cross sections at two
values of $\epsilon$.  The separated cross sections are determined at fixed
values of $W$, $Q^2$, common for both high and low values of $\epsilon$.
Because the acceptance covers a range in $W$ and $Q^2$, the measured cross
sections, and hence the separated response functions, represent an average over
this range.  They are determined at the average values (for both $\epsilon$
points together), $\overline{Q^2}$, $\overline W$, which are different for each
$t$ bin.  The experimental cross sections were calculated by comparing the
experimental yields to a Monte Carlo simulation of the experiment.  The
simulation uses a quasi-free $N(e,e'\pi^{\pm})N'$ model, where the struck
nucleon carries Fermi momentum, but the events are reconstructed in the same
manner as the experimental data, i.e. assuming the target is a nucleon at rest.
The Monte Carlo includes a detailed description of the spectrometers, multiple
scattering, ionization energy loss, pion decay, and radiative processes.

The separated cross sections, \sigl\ and \sigt, are shown in
Fig.~\ref{fig:xsec}.  Even if $\pi^+$ production on $^2$H occurs only on the
proton, the deuterium cross section cannot be directly connected to the free
$^1$H cross section because the Monte Carlo cross-section model ignores
off-shell effects and averages over the nucleon momentum distribution in $^2$H.
The uncertainties in the separated cross sections have both statistical and
systematic sources.  The statistical uncertainty in $\sigma_T+\epsilon
\sigma_L$ is 5-10\% for $\pi^-$ settings, and more uniformly near 5\% for
$\pi^+$ settings.  Systematic uncertainties that are uncorrelated between high
and low $\epsilon$ points are amplified by a factor of $1/\Delta \epsilon$ in
the $L$/$T$ separation.  This uncertainty ($\sim 1.3\%/\Delta\epsilon$) is
dominated by uncertainties in the spectrometer acceptance, uncertainties in the
efficiency corrections due to \u{C}erenkov trigger blocking and analysis cuts,
and the Monte Carlo model-dependence.  Scale systematic uncertainties of
$\sim$3\% (not shown in the figure) propagate directly into the separated cross
sections.  They are dominated by uncertainties in the radiative corrections,
pion decay and pion absorption corrections, and the tracking efficiencies.  The
systematic uncertainty due to the simulation model and the applied $M_X$ cut
(model-dependence) was estimated by extracting new sets of $L$/$T$/$LT$/$TT$
cross sections with alternate models and tighter $M_X$ cuts.

In the $\sigl$ response of Fig. \ref{fig:xsec}, the pion pole is evident by the
sharp rise at small $-t$.  $\pi^-$ and $\pi^+$ are similar, and the data at
different \qsq\ follow a nearly universal curve versus $t$, with only a weak
$Q^2$-dependence.  The $T$ responses are flatter versus $t$.

Finally, $\pi^-/\pi^+$ ratios of the separated cross sections were formed to
cancel nuclear binding and rescattering effects.  Many experimental
normalization factors cancel to a high degree in the ratio (acceptance, target
thickness, pion decay and absorption in the detectors, radiative corrections,
etc.).  The principal remaining uncorrelated systematic errors are in the
tracking inefficiencies, target boiling corrections, and \u{C}erenkov blocking
corrections.

\begin{figure}[h]
\begin{center}
\includegraphics[width=8.5cm]{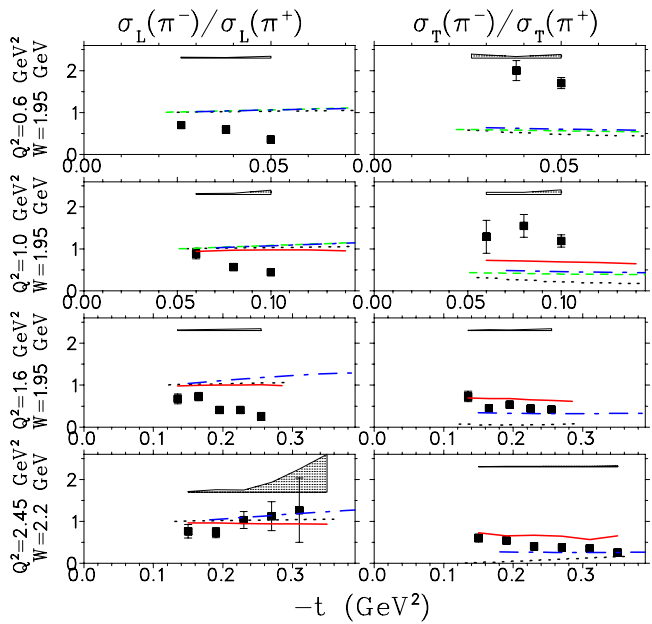}
\end{center}
\caption{(Color online) The ratios $R_L$ and $R_T$ versus
$-t$ for four \qsq\ settings.  The error bars include statistical and
uncorrelated systematic uncertainties.  The model-dependences of the ratios are
indicated by the shaded bands.  The dotted black curves are
predictions of the VGL Regge model \protect{\cite{van98}} using the values
$\Lambda_{\pi}^2=0.394$, 0.411, 0.455, 0.491 \gevsq, as determined from fits to
our $^1$H data \protect{\cite{Huber08}}, and the solid red curves are
predictions by Goloskokov and Kroll \protect{\cite{gk13}}, both models
calculated at the same $\overline{W}$, $\overline{Q^2}$ as the data.  The
dashed green curves are predictions by Kaskulov and Mosel
\protect{\cite{kaskulov}}, and the dot-dashed blue curves are the predictions
by Vrancx and Ryckebusch \protect{\cite{vrancx}}, both models calculated at the
nominal kinematics.}
\label{fig:Rlt_plot}
\end{figure}

Fig. \ref{fig:Rlt_plot} shows the first experimental determination of $R_L$.
The ratio is approximately 0.8 near $-t_{\rm min}$ at each \qsq\ setting, as
predicted in the large $N_c$ limit calculation of Ref. \cite{frankfurt}.  The
data are generally lower than the predictions of the pion-pole dominated models
\cite{van98,kaskulov, vrancx}.  Under the naive assumption that the isoscalar
and isovector amplitudes are real, $R_L=0.8$ gives $A_S/A_V=0.06$.  This is
relevant for the extraction of the pion form factor from electroproduction
data, which uses a model including some isoscalar background.  This result is
qualitatively in agreement with the findings of our pion form factor analyses
\cite{Huber08,volmer}, which found evidence of a small additional contribution
to $\sigl$ not taken into account by the VGL Regge Model in our $Q^2=0.6-1.6$
GeV$^2$ data at $W=1.95$ GeV, but little evidence for any additional
contributions in our $Q^2=1.6-2.45$ GeV$^2$ data at $W=2.2$ GeV.  The main
conclusion to be drawn is that pion exchange dominates the forward longitudinal
response even $\sim 10\ m_{\pi}^2$ away from the pion pole.

Also in Fig. \ref{fig:Rlt_plot} are the first $R_T$ results in
electroproduction.  At $Q^2$=0.6, 1.0 \gevsq, $R_T$ drops rapidly and given the
small $t$-range covered, it is not apparent if this drop is due to $t$ or
$Q^2$-dependence.  However, the values at $Q^2$=1.6 and 2.45 \gevsq\ overlap,
suggesting that $R_T$ is primarily a function of $-t$, dropping from about 0.6
at $-t=$0.15 to about 0.3 at $-t$=0.3 \gevsq.  Interestingly, photoproduction
data in this $t$-range \cite{heide} give simiar values.  It is noteworthy that
the unseparated data of Ref. \cite{Brauel1} reach a value of 0.3 at a much
higher value of $-t$.  A value of $-t$=0.3 \gevsq\ seems quite low for
quark-charge scaling arguments to apply directly.
This might indicate the partial cancellation of soft QCD corrections in the
transverse $\pi^-/\pi^+$ ratios.  Previous photoproduction
measurements of $R_T$ have hinted at quark-partonic behavior, but such
non-forward, $Q^2=0$ measurements are inherently more difficult to interpret
due to sea quark and $u$-channel contributions.  Indeed, the photoproduction
measurements at sufficiently high $-t$ first dip down toward 1/4 then {\em
increase} at backward angles \cite{Gao}.  The models of
Refs. \cite{VGL,kaskulov,vrancx} do not accurately predict $R_T$ at $-t_{\rm
min}$, although \cite{vrancx} does much better at higher $-t$.  The
Goloskokov-Kroll GPD-based model is in reasonable agreement, but the parameters
in this model are optimized for small skewness ($\xi<0.1$) and large $W>4$ GeV.
The application of this model to the kinematics of our data requires a
substantial extrapolation and one should be cautious in this comparison.
Indeed, although the model does a reasonable job at predicting the
$\pi^-/\pi^+$ ratios, the agreement of the model with $\sigt$ is not good
\cite{gk13}.  Further theoretical work is clearly needed to investigate
alternative explanations of the observed ratios.

To summarize, our data for $R_L$ trend toward unity at low $-t$, indicating the
dominance of isovector processes in forward kinematics, which is relevant for
the extraction of the pion form factor from electroproduction data
\cite{Huber08,volmer,hornt}.  The evolution of $R_T$ with $-t$ shows a rapid
fall off consistent with $s$-channel quark knockout.  Since $R_T$ is not
dominated by the pion pole term, this observable is likely to play an important
role in future transverse GPD programs.  Further work is planned after the
completion of the JLab 12 GeV upgrade, including complete separations at
$Q^2$=5-10 \gevsq\ over a larger range of $-t$ \cite{12gev}.

\begin{acknowledgments}
The authors thank Drs. Goloskokov and Kroll for the unpublished model
calculations at the kinematics of our experiment, and Drs. Guidal, Laget, and
Vanderhaeghen for modifying their computer program for our needs.  This work is
supported by DOE and NSF (USA), NSERC (Canada), FOM (Netherlands), NATO, and
NRF (Rep. of Korea).  Additional support from Jefferson Science Associates and
the University of Regina is gratefully acknowledged.  At the time these data
were taken, the Southeastern Universities Research Association (SURA) operated
the Thomas Jefferson National Accelerator Facility for the United States
Department of Energy under contract DE-AC05-84150.

\end{acknowledgments}

\end{document}